# Intrinsic DNA Curvature of Double-Crossover Tiles

**Seungjae Kim,[1] Junghoon Kim,[1] Pengfei Qian,[2] Jihoon Shin,[3] Rashid Amin,[3] Sang Jung Ahn,[4] Thomas H. LaBean,[5] Moon Ki Kim,[2,3]* and Sung Ha Park,[1,3]***

[1]Department of Physics, Sungkyunkwan University, Suwon 440-746, Korea,

[2]School of Mechanical Engineering, Sungkyunkwan University, Suwon 440-746, Korea,

[3]Sungkyunkwan Advanced Institute of Nanotechnology (SAINT), Sungkyunkwan University, Suwon 440-746, Korea,

[4]Korea Research Institute of Standards and Science, Daejeon 305-340, Korea,

[5]Departments of Computer Science and Chemistry, Duke University, Durham, NC 27708, USA

*Corresponding Authors

E-mails: mkkim1212@skku.edu (MKK) and sunghapark@skku.edu (SHP)

**Abstract.** A theoretical model which takes into account the structural distortion of double-crossover DNA tiles has been studied to investigate its effect on lattice formation sizes. It has been found that a single vector appropriately describes the curvature of the tiles, of which a higher magnitude hinders lattice growth. In conjunction with these calculations, normal mode analysis reveals that tiles with relative higher frequencies have an analogous effect. All the theoretical results are shown to be in good agreement with experimental data.

With the advent of structural DNA nanotechnology, much effort has gone into the fabrication of novel DNA nanostructures [1-14]. Although major advances have taken place during the last decade, some fundamental issues such as symmetry arguments and improving DNA crystal resolutions still remain to be resolved [15]. At the core of the issues lies a lack of understanding of the exact atomic scale structural properties of the basic building blocks which form the basis of DNA nanostructures. Of the many designed building blocks created to date, one of the most widely known and used is the double-crossover (DX) tile [16]. The sheer number of different types of structures, e.g., ribbons [17, 18], tubes [19, 20], and lattices [3, 21, 22], that have been fabricated using DX tiles as a motif is a testament to its popularity. Here we present a detailed analysis of the structural properties of the DX tile with its intrinsic curvature and its effect on DX lattice formations based on structural data of DNA sequences taken from the Nucleic Acid Database (NDB) [23]. If DNA nanotechnology is to make further progress in becoming a viable bottom-up solution to the limitations of the top-down fabrication approach, the intrinsic curvature of various DNA motifs should be taken into account in future works of DNA lattice fabrication and perhaps even exploited.

The basic components of a DX tile consist of four DNA strands forming two duplexes which are connected by two crossover junctions. DX tiles having different sequences, and thus slightly different structures, can be constructed by exploiting the inherent programmability of DNA. In this manner, four different types of DX tiles were designed for this study as shown in Figure 1. Each DX tile was fabricated from four DNA strands comprising two long and two short strands. The short strands have elongated sticky ends which act as adhesive bindings with other complementary DX tiles. The tiles were designed so that the sticky ends labeled c1’, c2, c3’, and c4 of DX1 would bind with c1, c2’, c3, and c4’ of DX2, respectively, to form 2D lattices consisting of two-tile units of DX1 and DX2. Analogously, c1’, c2, c5’, and c6 of DX3 bind with c1, c2’, c5, and c6’ of DX4, respectively.

The intrinsic curvature of the DX tile originates from the sequence dependence of the DNA flexibility [24, 25]. Figure 2 represents a schematic diagram of a single DX tile with its curvature. In order to properly describe the curvature, we consider not only sequence dependent fluctuations but also non-neighbor interactions between the base pairs, whereby units of tetramers (four base pairs) were considered, in accordance with Ref. 26. Following in the footsteps of Ref. 27, four different structural parameters - twist, roll, tilt, and slide - of the duplex DNA molecules that characterize the orientation of the plane made by successive base pairs were used. The origin, **O**, the DX tile is taken to be the midpoint of the line connecting the $C_8$ of a purine (R) and $C_6$ of a pyrimidine (Y), i.e. $RC_8{-}YC_6$, of the first base pair of the lower duplex. The unit DX tile is always taken to be from $n = 1$ to $n = 37$, where $n$ represents the base pair index, as shown in Figure 2b. Several vectors needed to explain the curvature, e.g., $\mathbf{A}_n$, $\mathbf{B}$, $\mathbf{C}_n$, and $\mathbf{D}_n$, have been defined and are depicted in Figure 2. Of these, the one pivotally related to the curvature is $\mathbf{D}_n$, which we shall refer to as the curvature displacement vector and defined as the difference between $\mathbf{C}_n$ and $\mathbf{A}_n + \mathbf{B}$, i.e., $\mathbf{D}_n = \mathbf{C}_n - (\mathbf{A}_n + \mathbf{B})$. The magnitude of $\mathbf{D}_n$, is a direct measure of the degree of curvature up to the $n^{th}$ base pair.

Figure 3 shows the actual results of our calculations. The six graphs are plots of the magnitude of the D-vector (y-axis) versus the distance between the the first base pair ($n = 1$) and the $n^{th}$ base pair (x-axis), where $1 \le n \le 37$. Figure 3a, b, d, and e are plots of $|\mathbf{D}_n|$ for single DX tiles and Figure 3c and f are plots of two-tile systems, DX1+DX2 and DX3+DX4, respectively, which are the actual building blocks of the lattice. Henceforth, we shall refer to lattices formed from DX1 and DX2 as lattice A and lattices formed from DX3 and DX4 as lattice B. Since the length of one DX tile is 37 base pairs and one helical full turn of the double helix is ~10.5 base pairs, a single DX tile makes ~3.5 full turns. In order for two tiles to bind, the half turn of the first tile type must be accounted for and the second tile type must rotate around the $\mathbf{A}_f$ axis by 180 ° to bind with the first tile type. The plots reveal that the DX tiles have varying degrees of curvature, but the most important feature of a DX tile regarding lattice formation are captured in the positions of the four *end* points which are the midpoints of the $RC_8{-}YC_6$ line of the first and last base pairs of the upper and lower duplexes. It is near these end points that the bindings between tiles occur. If the magnitude of $\mathbf{D}_f$ is non-zero, then these four end points do not all lie on a parallelogram and due to geometrical constraints, the binding of DX tiles is hindered and the formation of a 2D lattice is somewhat suppressed. In the *ideal* case where $\mathbf{D}_f$ is zero, the four end points all lie on a parallelogram and the geometrical constraints for the formation of a 2D lattice are lifted, rendering conditions conducive to larger lattices.

To check the curvature dependence on the lattice formation size, analyses of DNA lattices were carried out by atomic force microscopy (AFM). Figure 4a and b show images of lattices A and B, respectively. The regions enclosed in red boundaries indicate formations of typical crystals. The area of 12 crystals were analyzed and found to have an average value of ~$1.26 \times 10^5$ $nm^2$ and ~$1.04 \times 10^5$ $nm^2$ for lattices A and B, respectively. Since the size of single DX tile is about 50 $nm^2$, the average number of DX tiles per crystal is ~2500 and ~2000 for each lattice type. This trend is in good agreement with our curvature displacement vector calculations. From Figure 4c, we see that the magnitude of $\mathbf{D}_f$ is roughly 50% larger for lattice B (~7.6 Å) than for lattice A (~5.2 Å), impeding the growth of lattice B compared to lattice A.

Another contributing factor in lattice growth may be due to the differences in the vibrational modes of the DX tiles. Normal mode analyses (NMA) [28, 29] within the framework of the elastic network model (ENM) [30, 31], whereby the potential of the atoms in the molecule is approximated as a Hookean harmonic function, were performed for the four single DX tiles as well as the two-tile units which make up the lattice. A summary of the four lowest vibrational modes of the two-tile units are presented in Figure 5, where it can be seen that for both two-tile systems, the lowest vibrational mode (1st mode), i.e., the most dominant vibrational mode, corresponds to the bending motion out of the plane where the duplexes of the DX tiles lie. Also, for the four vibrational modes shown, the frequencies of the vibrations for the DX1+DX2 unit system were lower than that of the DX3+DX4 unit system. The lower frequencies can be attributed to a lower stiffness since, due to the Hookean potential, the frequency is proportional to the square root of the stiffness over the mass of the system (the molecular weights of the two systems are within ~0.16% of each other). This decrease in stiffness of the DX1+DX2 system implies a more flexible unit in which the bindings between the tiles may form more easily leading to larger lattices.

In conclusion, the effects of the intrinsic curvature and vibrational modes of DX tiles on lattice growth have been studied. The magnitude of a curvature displacement vector, |**D**|, suffices in representing the curvature of DX tiles and is experimentally shown to be an appropriate indicator of relative lattice sizes. NMA results also suggest that higher frequencies of the same vibrational modes may suppress larger lattice growth. Although the focus of the study presented here has been on lattice formations of a particular type of DX tiles, i.e. double anti-parallel odd (DAO) type DX tiles, further theoretical analyses on double anti-parallel even (DAE) type DX tiles [32], triple-crossover tiles [5], and even origami structures [11,33,34] can be straightforwardly carried out using the same strategy and will inevitably lead to a deeper understanding of their structural and dynamical characteristics, expediting the development of new motifs with higher production yields.

**Acknowledgements** MKK was supported by the World Class University Program through the National Research Foundation of Korea (NRF) funded by the Ministry of Education, Science and Technology (R33-10079) and SHP by the Basic Science Research Program through the National Research Foundation of Korea (NRF) funded by the Ministry of Education, Science and Technology (2010-0013294).

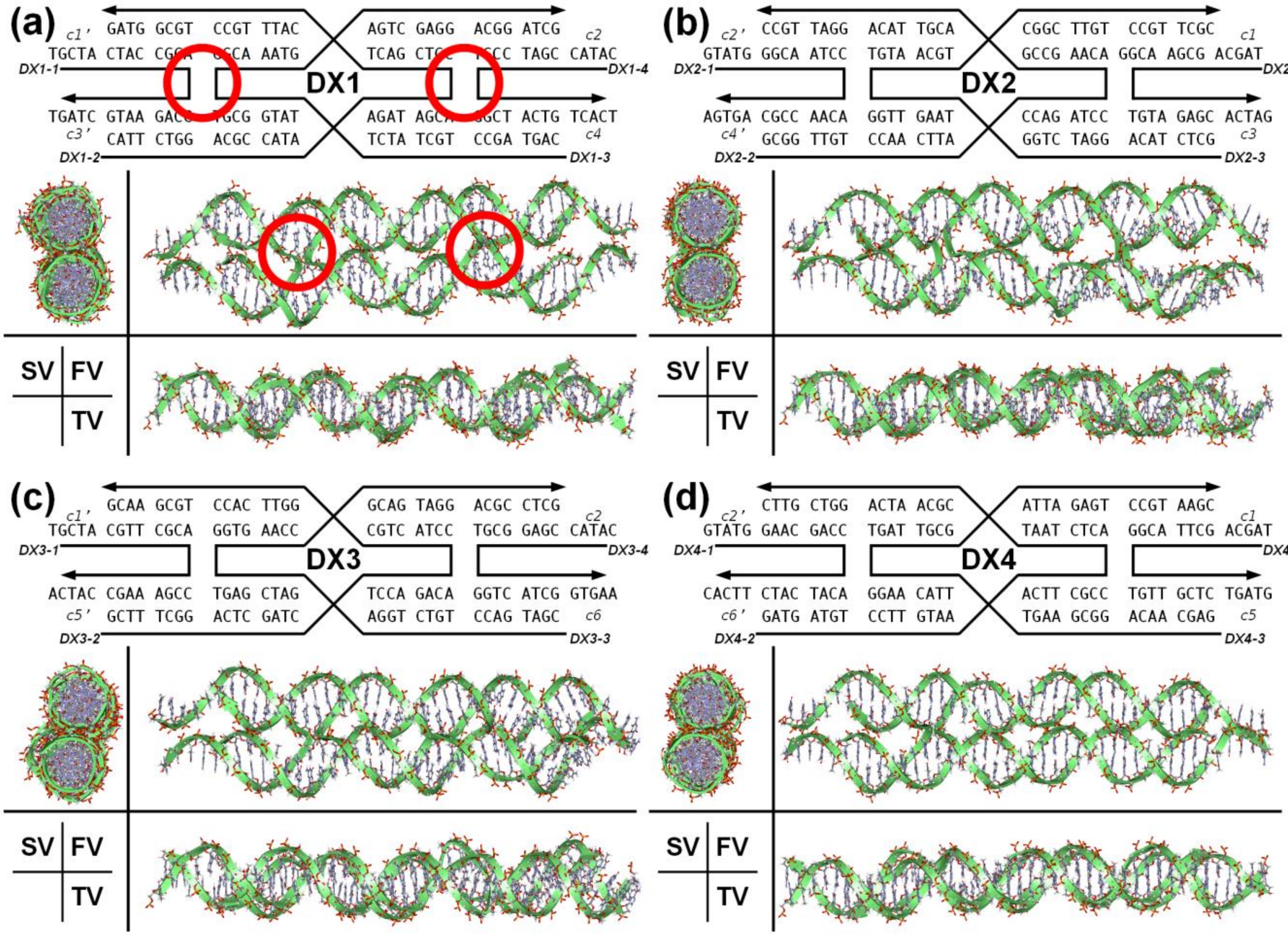

**Figure 1.** DNA sequence maps and molecular visualizations of the four types of DX tiles: (a) DX1, (b) DX2, (c) DX3, and (d) DX4. Three different perspectives are given for each DX tile, Front View (FV), Top View (TV), and Side View (SV). The green ribbons represent the backbones running through the nucleotides of each strand with the arrows indicating the 5' to 3' direction. The red circles show positions of crossover junctions.

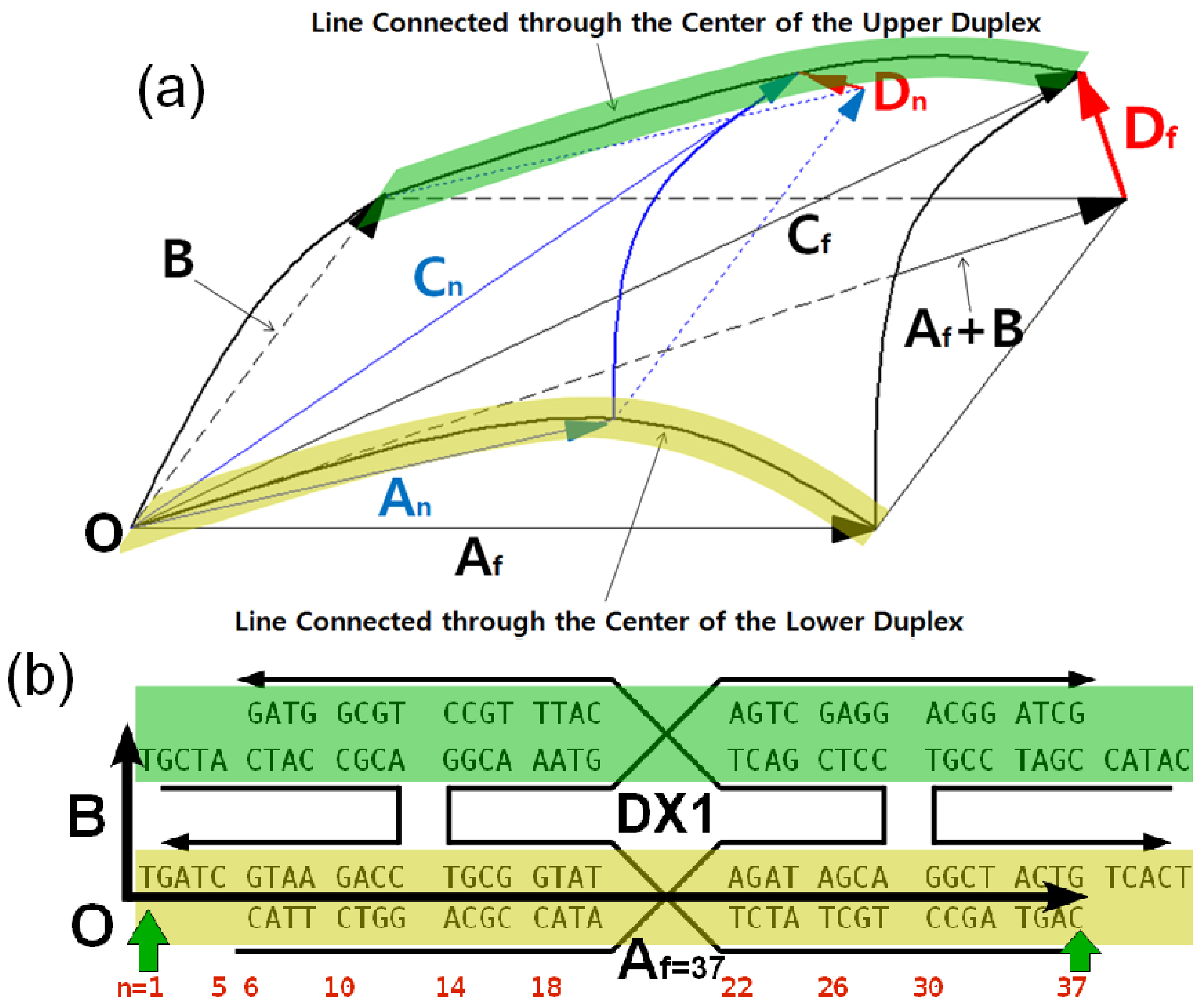

**Figure 2.** A schematic diagram illustrating the vectors defined for use in the calculations. (a) Shaded green and yellow areas represent the axes which run through the center of the upper and lower duplexes, respectively. The origin, **O**, of the DX tile is taken to be the midpoint of the $RC_8$-$YC_6$ line of the first base pair of the lower duplex. $\mathbf{A}_n$ is a displacement vector which starts from the midpoint of the first base pair and ends at the $n^{th}$ base pair of the lower duplex. **B** is a constant vector defined as the difference between the midpoint of the first base pair of the upper duplex and the origin. $\mathbf{C}_n$ is the difference between the midpoint of the $n^{th}$ base pair and the origin. A curvature displacement vector, $\mathbf{D}_n$ is defined as $\mathbf{D}_n = \mathbf{C}_n - (\mathbf{A}_n + \mathbf{B})$, where $n$ runs from $1 \leq n \leq 37$ (= $f$ ) for single DX tile. (b) Corresponding diagram of a DX tile.

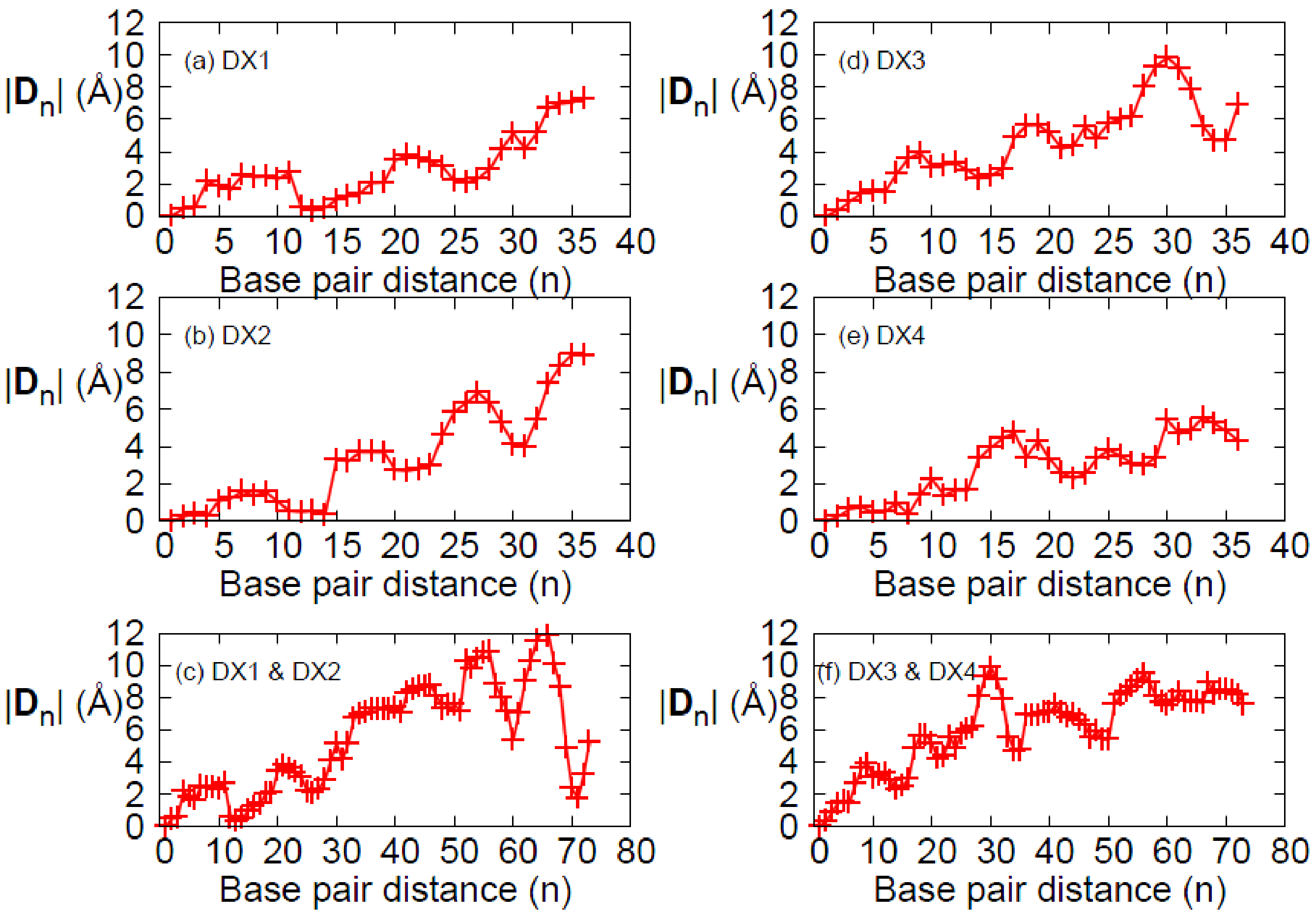

**Figure 3.** The magnitude of **D** for (a) DX1, (b) DX2, (c) unit tile of DX1 and DX2, (d) DX3, (e) DX4, and (f) unit tile of DX3 and DX4 with respect to the base pair distance. The base pair distance is taken as the distance between the first base pair ($n = 1$) and the $n^{th}$ base pair, where $1 \leq n \leq 37$.

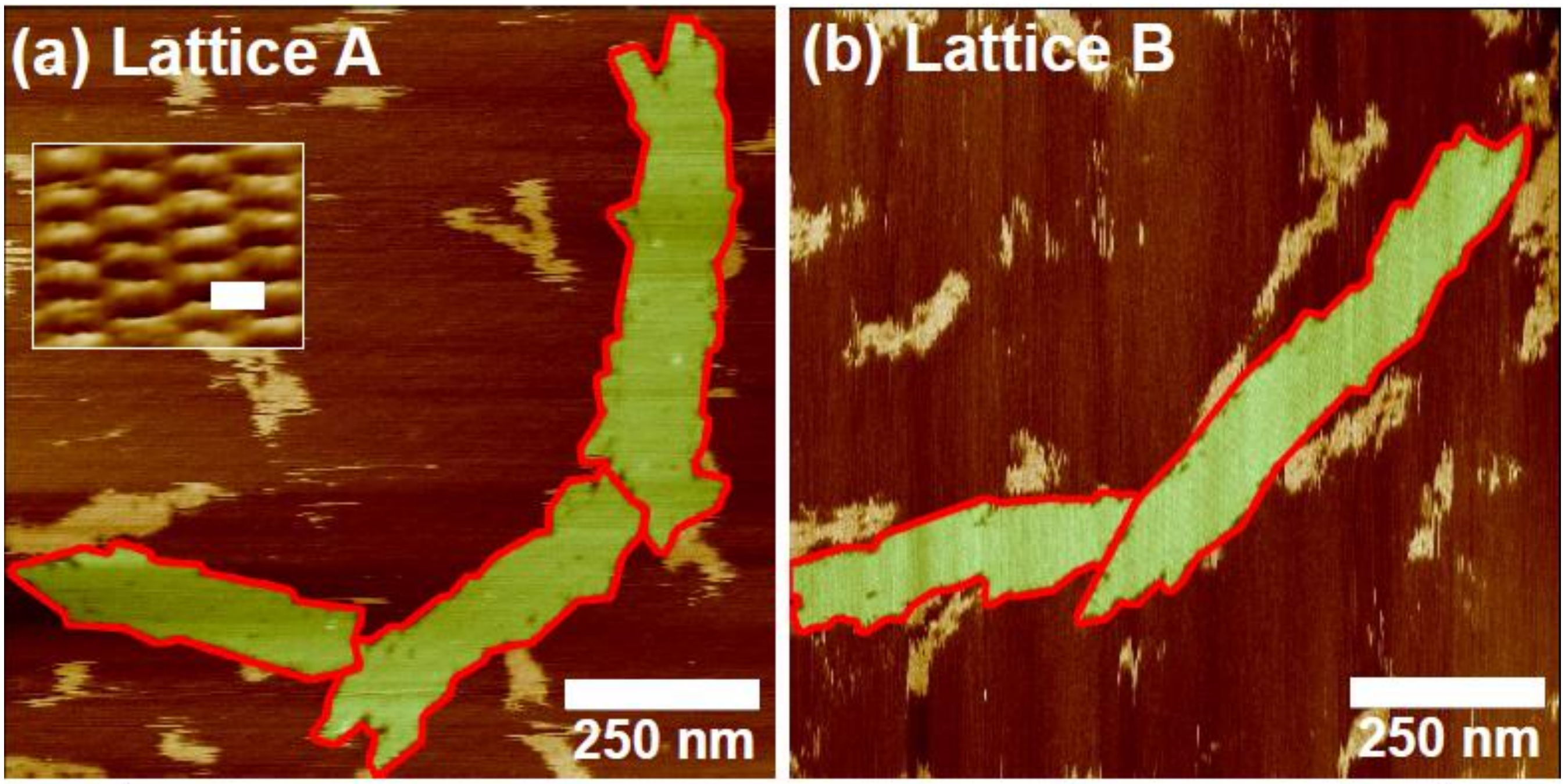

(c)

| | Lattice A | Lattice B |
|---|---|---|
| $\lvert \mathbf{D}_f \rvert$ (Å) | 5.24 | 7.57 |
| area/crystal ($nm^2$) | ~ 1.26 X $10^5$ | ~ 1.04 X $10^5$ |
| area/tile ($nm^2$) | 50.32 | 50.32 |
| number of tiles/crystal | ~ 2500 | ~ 2000 |

**Figure 4.** AFM images of (a) lattice A and (b) lattice B. The regions enclosed by red boundaries indicate single lattice formations. Individual DX tiles can be seen through the high resolution AFM image in inset (a), scale bar; 10 nm. (c) Number of tiles per crystal with respect to $\lvert \mathbf{D}_f \rvert$. Smaller values of $\lvert \mathbf{D}_f \rvert$ engender larger lattices.

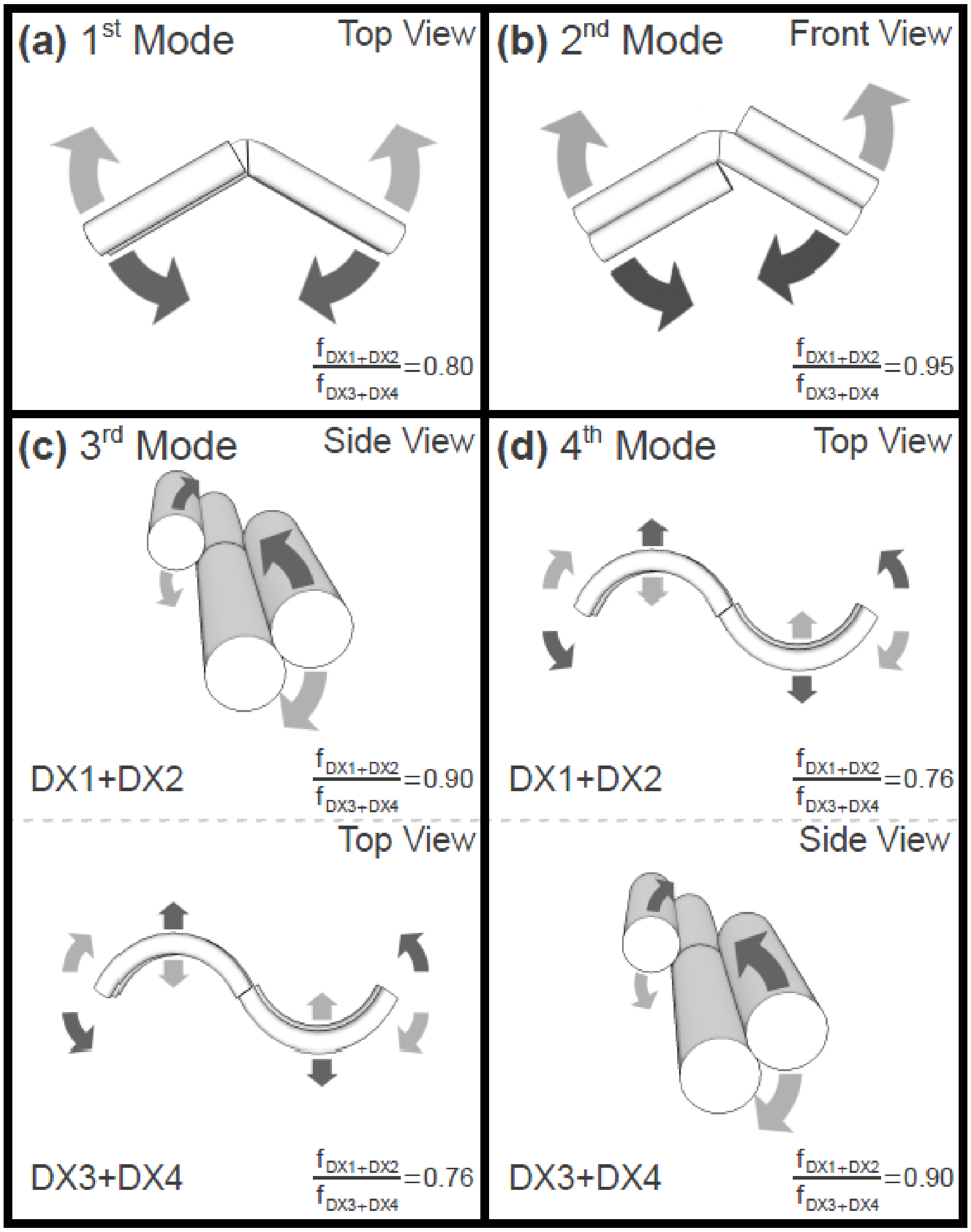

**Figure 5.** A schematic diagram of the four lowest modes (a)-(d) of the two-tile units obtained from NMA. The DX tiles are illustrated as simplified duplexes to clarify the vibrational modes of the two-tile units. Lower modes correspond to states with higher probabilities. Arrows of the same grayscale shading represent in-phase displacements. With the exception of the 3$^{rd}$ and 4$^{th}$ modes (c) and (d), the types of vibrations are the same for both (DX1+DX2) and (DX3+DX4) units; (a) out-of-plane bending motions, (b) in-plane bending motions, (c) twisting motions (DX1+DX2) and out-of-plane bending motions (DX3+DX4), and (d) out-of-plane bending motions (DX1+DX2) and twisting motions

(DX3+DX4). The relative frequencies between the two two-tile units, $(f_{DX1+DX2})/(f_{DX3+DX4})$ are always less than unity, strongly suggesting that the (DX1+DX2) unit is more flexible than (DX3+DX4).